\documentclass[12pt]{article}

\usepackage{scicite}
\usepackage{graphicx}
\usepackage[colorlinks]{hyperref}
\usepackage{amsmath}

\usepackage{times}

\newenvironment{sciabstract}{%
\begin{quote} \bf}
{\end{quote}}

\title {Wave Energy Is Conserved in a Spatially Varying and Inhomogeneously Moving Medium }

\author
{Zhaohua Wu,$^{1,2\ast}$ Jie Sun,$^{1}$ Zhe-Min Tan,$^{3}$ Ming Cai,$^{1}$ Yongyun Hu,$^{4}$ \\
	Norden E Huang$^{5}$\\
	\\
	\normalsize{$^{1}$Department of Earth, Ocean, and Atmospheric Science, Florida State University,}\\
	\normalsize{Tallahassee, FL 32306, USA}\\
	\normalsize{$^{2}$Center for Ocean-Atmospheric Prediction Studies, Florida State University,}\\ 
	\normalsize{Tallahassee, FL 32306, USA}\\
	\normalsize{$^{3}$School of Atmospheric Sciences, Nanjing University,}\\ 
	\normalsize{Nanjing, China}\\
	\normalsize{$^{4}$Dept. of Atmospheric and Oceanic Sciences, Peking University,}\\ 
	\normalsize{Beijing, China}\\
	\normalsize{$^{5}$First Institute of Oceanography, Qingdao, China}\\ 
	\\
	\normalsize{$^\ast$To whom correspondence should be addressed; E-mail:  zwu@fsu.edu.}
}

\date{}

\begin{document} 


\baselineskip24pt


\maketitle 

\pagebreak


\begin{sciabstract}
 
 Waves are propagating disturbances that redistribute energy across space ({\it 1--4}). Previous studies ({\it 5--7}) have shown that for waves propagating through an inhomogeneously moving mean flow, the conserved quantity is wave action rather than wave energy, raising questions about the validity of energy conservation—one of the foundational principles of physics. In this study, we prove that wave action conservation is, in fact, an apparent form of wave energy conservation in spatially varying and inhomogeneously moving media, where waves undergo deformation during propagation. We further show that wave action conservation can be derived directly from the law of energy conservation. This result holds universally across all isolated wave systems in varying media, including hydrodynamic and non-hydrodynamic waves.
  
\end{sciabstract}

\pagebreak

\section*{Introduction}
Hydrodynamic waves are propagating disturbances that travel through the continuous adjustment of unbalanced states between neighboring elements. This process involves the transformation between kinetic and potential energy({\it 8}). Kinetic energy is typically associated with the inertial properties of the wave system, while potential energy corresponds to the elastic properties that characterize the medium. In this context, a medium is not merely material but also encompasses the dynamic properties it possesses, which define its behavior and interaction with the propagating waves.

For hydrodynamic waves, it is generally believed that the governing conservation law is that of wave action rather than wave energy. The concept of wave action conservation was first derived approximately sixty years ago by Whitham ({\it 5}), who applied the variational principle to a set of Boussinesq equations. Whitham demonstrated that the ratio of the local wave energy density to the wave frequency, which he referred to as the "adiabatic invariant," remains constant even though the energy density itself varies during wave propagation. Subsequently, Whitham's work was expanded upon ({\it 6, 8}). The "adiabatic invariant" was later renamed "wave action" ({\it 8}) to align with Lagrange’s definition of action ({\it 10}). Wave action conservation can also be derived using the WKB approximation and the stress-energy tensor ({\it 7, 11}).

However, a fundamental question remains: why does the energy itself not remain conserved during wave propagation? The energy conservation has been a cornerstone of physics for nearly two centuries, serving as one of its foundational principles. Any evidence suggesting the non-conservation of energy in an isolated physical system would challenge this fundamental principle, indicating potential flaws in the current framework of physics. Such claims must be rigorously examined and reconciled to maintain the coherence of the scientific understanding of natural laws.

Efforts to reconcile wave action conservation with energy conservation have accompanied studies of the wave action law since its inception. Bretherton and Garrett ({\it 6}) devoted a substantial section to discussing why wave energy is not conserved, emphasizing the potential for energy exchange between waves and the medium. They highlighted the challenge of separating the varying mean flow from the waves superimposed on it and minimizing their interaction—particularly for hydrodynamic waves in an inhomogeneously moving medium. The idea that wave energy can be exchanged with the medium through processes such as reflection, scattering, refraction, and absorption in a varying environment has been widely accepted to reconcile wave energy conservation in various systems, including sound waves ({\it 12}), plasma waves ({\it 13}), and seismic waves ({\it 14}). However, a more fundamental confirmation of energy conservation across all wave systems would be a derivation that begins with the energy conservation law and leads directly to wave action conservation—without invoking any exchange of energy between the wave and its medium.

This study seeks to address the reconciliation challenge precisely in the manner proposed above. We consider the general case of waves in a varying medium, treating the case of an inhomogeneously moving medium as a special case. We show that the apparent wave action conservation is, in fact, a manifestation of the underlying energy conservation law. This reconciliation unifies the governing conservation principles for wave propagation across diverse systems, providing a consistent framework that reaffirms the fundamental role of energy conservation in physics. Since the entire proof does not require any specific type of wave, energy conservation remains valid not only for hydrodynamic waves but also for non-hydrodynamic waves, such as electromagnetic waves propagating in a varying medium.

\section*{From Wave Action Conservation to Wave Energy Conservation}

The process of reconciliation is schematically illustrated in Fig. 1. It consists of two parts: (1) transforming the traditional wave action, which is a mathematically derived quantity for the special case of an inhomogeneously moving constant medium, into a more general form with clearer physical meaning that applies to any continuously varying medium; and (2) deriving the general form of wave action conservation from a general waveform and the energy conservation law. The detailed mathematical proof can be found in the Supplementary Material. In the following, we discuss the key steps and their physical interpretations.

\begin{figure}[!h]
	\centering
	\includegraphics[width=0.9\linewidth]{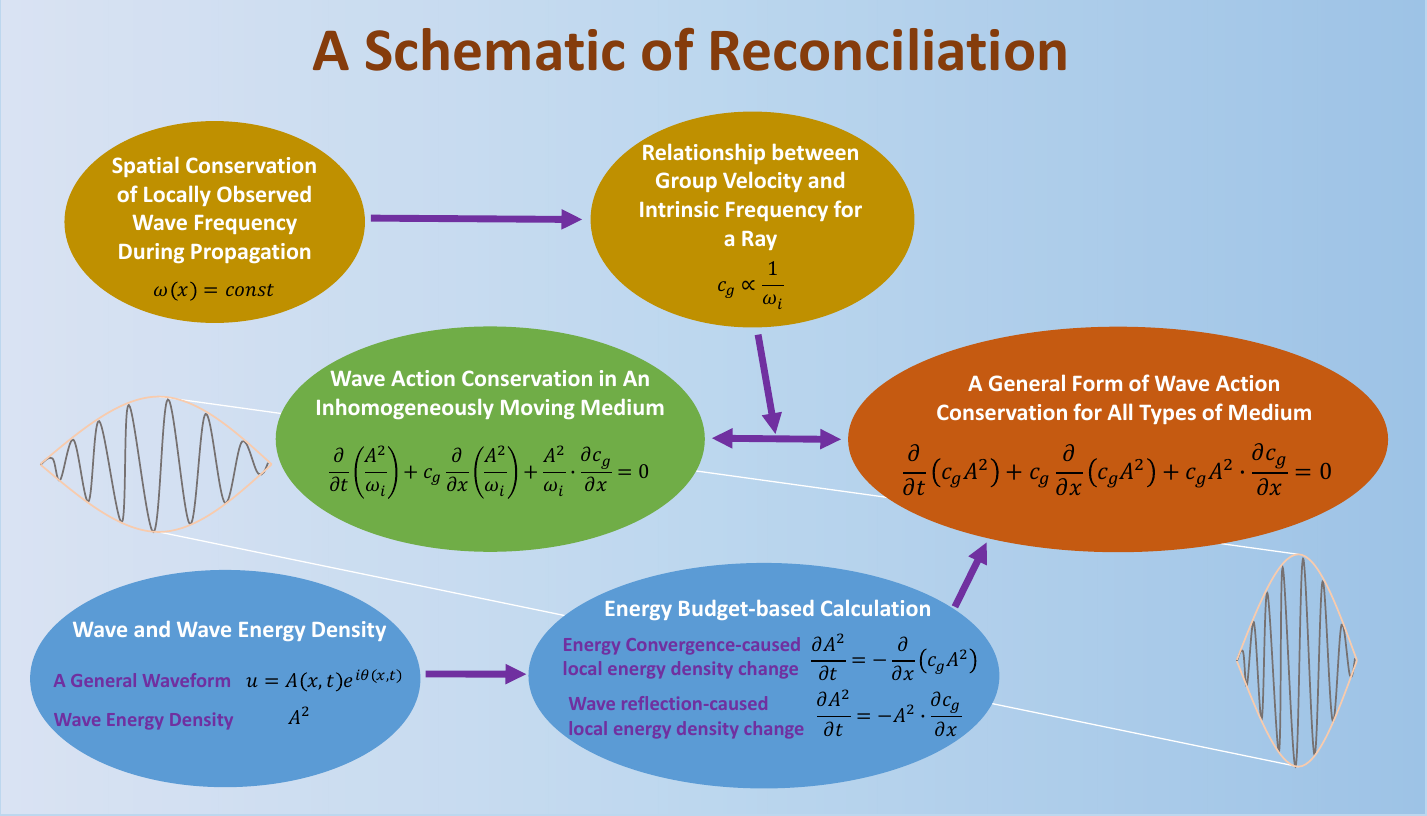}
	\caption{A schematic diagram illustrating the reconciliation between wave energy and wave action conservation. The background of the figure represents a varying continuous medium, with the horizontal axis representing space and the vertical axis representing time, and a wave packet propagating in the medium. The darker-colored region (on the right-hand-side) corresponds to a medium with a slower wave packet propagation speed. 
	}
	\label{Figure_1}
\end{figure}

Two key concepts are involved in the derivation of wave action conservation: the wave packet, which is a spatially localized bundle of waves, and ray theory, which focuses on the propagation path of a wave packet. In most previously considered cases, it is assumed that the packet contains little spread of frequency, making it almost non-dispersive. Although a dispersive case can also be considered, in such a case, the packet's coherence will not be preserved.

In this case, the wave packet can be expressed as  $A(x,t)e^{i\theta (x,t)}$, where $A(x,t)$ is the amplitude of the wave bundle and and its square is the wave energy density. $\theta (x,t)$ is the phase function, whose (negative) temporal change rate gives locally observed frequency $\omega$ and spatial change rate gives wavenumber $k$. For simplicity, here we first consider a one spatial dimensional case. 

When the packet propagates in an inhomogeneously moving constant medium, the intrinsic frequency, $\omega _i$, which is the wave frequency measured by an observer moving at the local speed of the medium, also needs to be considered. These two frequencies are related by the Doppler shift caused by the medium's movement, i.e., $\omega (x) = \omega _i(x) + U(x)k(x)$, where $U(x)$ is the spatially varying medium moving speed. The wave action density is defined as $A^2/\omega_i$ and its conservation is expressed as
\begin{equation}
	\frac {\partial} {\partial t} \left ( \frac {A^2} {\omega _i} \right ) + c_g \frac {\partial} {\partial x} \left ( \frac {A^2} {\omega _i} \right )  + \frac {A^2} {\omega _i}  \cdot \frac {\partial c_g} {\partial x}=0.
	\label{waveActionConservation}
\end{equation}
Equation (1) has the same form as the continuity equation for mass in fluid dynamics when the wave action density is replaced by mass density and packet propagation velocity, $c_g$, replaced by the mass moving speed. This is not surprise since the density of any conservation quantity should satisfy the same expression.

Our first step (Supplementary Material Section S.2) toward reconciliation is to prove that the locally observed frequency, $\omega$, at any spatial location, $x$, remains constant for a non-dispersive wave as it propagates forward, provided that the location is tracked following the wave propagation speed. Physically, this is intuitive, as an oscillating point in the wave acts as a disturbing force for the neighboring point in the direction of wave propagation, causing the neighboring point to inherit the same frequency of oscillation. Our proof is geometrically based and applies to all wave systems, in contrast to the WKB method, which typically begins with a specific wave system and requires additional assumptions.  

Our second step (Supplementary Material Section S.3) is to transform the original wave action density, $A^2/\omega_i$, into an alternative form, $c_gA^2$. With this transformation, the wave action conservation equation becomes
\begin{equation}
	\frac {\partial} {\partial t} ( c_gA^2 ) + c_g \frac {\partial} {\partial x}  ( c_gA^2 )  + c_gA^2  \cdot \frac {\partial c_g} {\partial x}=0.
	\label{alternativeWaveActionConservation}
\end{equation}
This transformation leverages the properties of a wave packet in a varying medium, the constancy of the locally observed frequency $\omega$ in a propagating wave, and the Doppler shift formula to infer that the intrinsic frequency $\omega_i$ is inversely proportional to the group velocity $c_g$. Since the wavenumber $k$ for such a wave is inversely proportional to the group velocity [Eq. (S7)], the group velocity $c_g$ in the quantity $c_gA^2$ becomes a parameter that characterizes the spatial scale of a wave packet and thus represents the shrinking or swelling of the wave packet or a fraction of it. Consequently, the alternative form of the wave action density, $c_gA^2$, corresponds to the wave energy density after accounting for the shrinking/swelling effect. In this sense, wave action conservation can be interpreted as wave energy conservation for the entire wave packet. 

\section*{From Wave Energy Conservation to Wave Action Conservation}    

Strictly speaking, the derivation in the last section relies on ray theory, in which the wave packet is non-dispersive, meaning that the wave envelope retains its shape except for proportional spatial shrinking or swelling during propagation. In the following, we derive wave action conservation [Eq. (2)] starting from wave energy conservation. Notably, this derivation does not require the assumption of non-dispersive waves.

\begin{figure}[!h]
	\centering
	\includegraphics[width=0.9\linewidth]{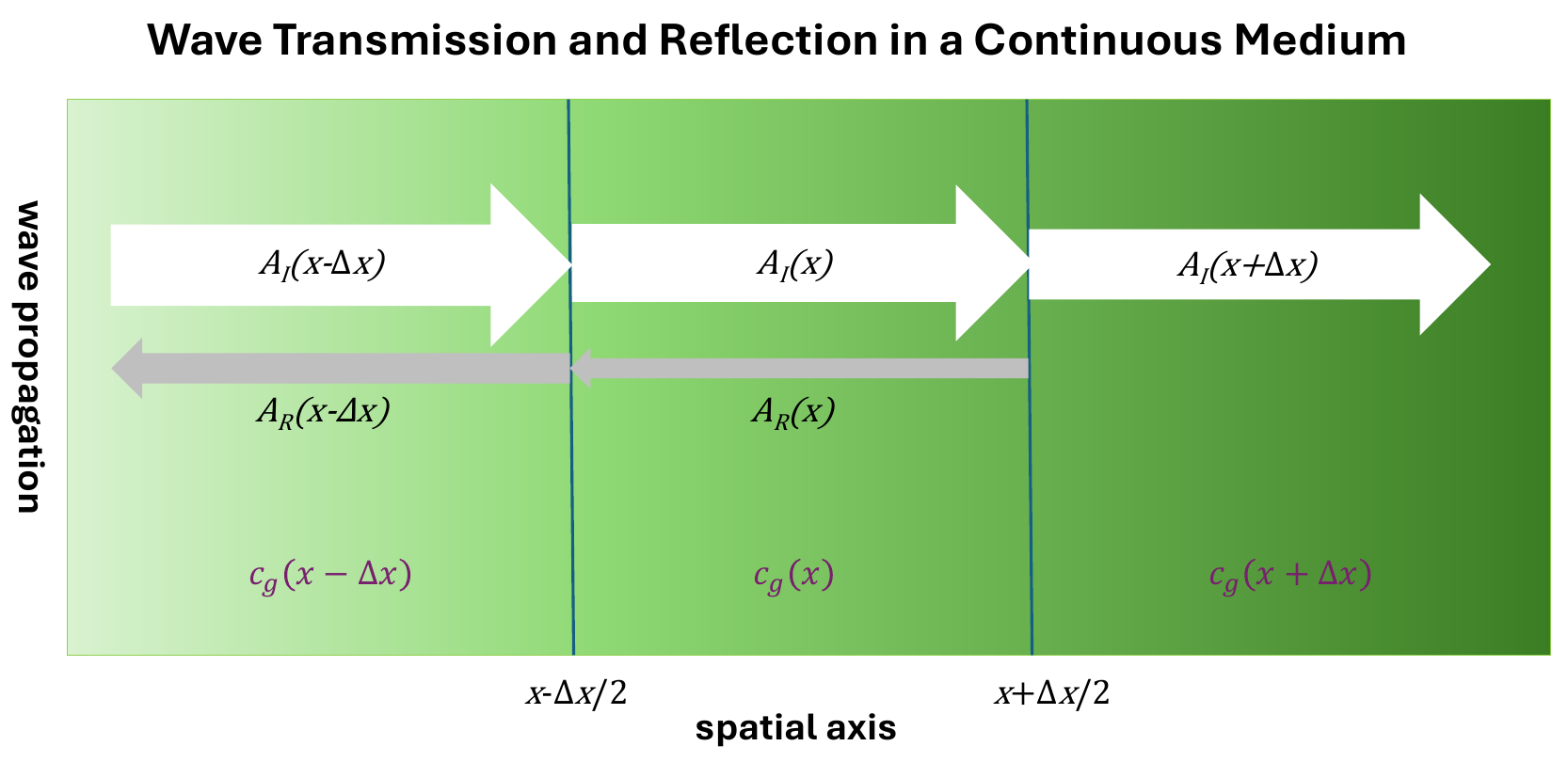}
	\caption{A schematic diagram of wave transmission and reflection in a continuously varying medium. A continuously varying medium is approximated by piece-wise constant media in the spatial direction. The white arrows represent the forward wave and gray arrows the reflected waves. 
	}
	\label{Figure_2}
\end{figure}  

This derivation utilizes the discrete representation of a continuous medium, similar to how integration is derived for a continuous function. The medium is divided into infinitesimal segments of width $\Delta x \to 0$, (see Fig. 2). When discontinuities are present, part of the propagating wave energy is reflected at the interface, satisfying 
\begin{equation}
	\Delta {A^2} = -\frac{\Delta c_g}{c_{g}} A^2.
	\label{reflectedWaveEnergy}
\end{equation}
Since the reflected wave energy cannot be accounted for as part of the incident wave energy within an infinitely small segment from $x-\Delta x/2$ to $x+\Delta x/2$, the net wave energy gain is given by the energy influx $c_g(x)A^2(x)$ minus the sum of the energy outflux $c_g(x+\Delta x)A^2(x+\Delta x)$ and the reflected portion $-(\Delta c_g/c_g) A^2$. The law of energy conservation dictates that this local net energy gain must result in a temporal change in the local energy density over the entire spatial segment. Since the time $\Delta t$ taken by the propagating wave to traverse the segment is $\Delta x /c_g$, we can directly obtain Eq. (2), the alternative form of the wave action conservation, from this reasoning. A more detailed exposition can be found in Supplementary Material Sections S.4–S.6.

\begin{figure}[!h]
	\centering
	\includegraphics[width=0.9\linewidth]{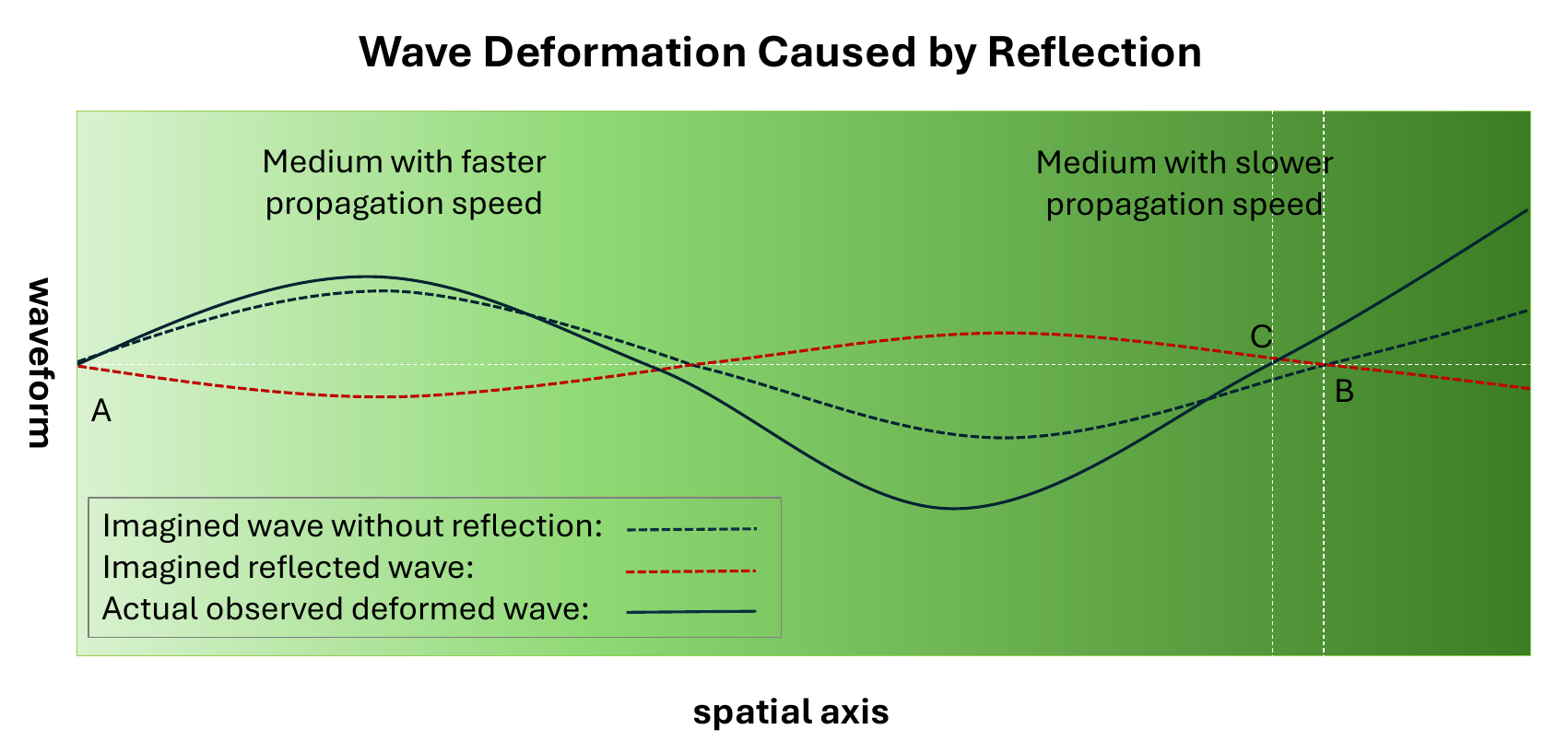}
	\caption{A schematic diagram illustrating wave deformation due to reflection. The meanings of the curves are provided in the legend. Note that the amplitudes of different waves are not proportional.
	}
	\label{Figure_3}
\end{figure}  

In this part of the derivation, it is shown that the reflected portion contributes to the deformation of the waveform in a continuous medium. This reflection is not considered an additional wave energy source or sink during wave propagation; rather, it is interpreted as a mechanism for waveform deformation through continuous reflections of infinitesimal amplitude as the wave propagates in a continuous medium. This is intuitively understandable in a medium with decreasing wave propagation speed, as illustrated in Fig. 3. Consider an imagined incident wave with upward zero-crossing points at A and B in the absence of reflection. Suppose an observer moves with point A. In a medium where propagation speed decreases, the reflected portion of the wave undergoes a $180^o$ phase shift. If we sum these two waves, the observer will observe the next upward zero-crossing point at C (solid line), leading to a shorter wavelength and an increase in wave amplitude. Since the actual wave under discussion is represented by the solid line, the reflected portion is inherently part of the propagating wave. In this sense, the reflected wave does not indicate a loss of wave energy but rather contributes to waveform deformation. A similar argument applies to the case of a medium with increasing propagation speed, except that the reflected wave undergoes no phase shift.

\section*{Summary and Discussion}

In this study, we begin with a general waveform that incorporates spatiotemporal variations in both amplitude and wavenumber within a varying medium. We demonstrate that the so-called wave action conservation is, in fact, a manifestation of wave energy conservation. Unlike previous studies that introduced external energy sources or sinks ({\it 12–14}), our reconciliation relies solely on a reinterpretation of the role of reflected waves during propagation in a varying medium. No explicit energy exchange with the medium is required; instead, the medium is treated simply as the material through which the wave propagates.

Our conclusion applies to all wave types, making it both general and robust, regardless of whether the wave under consideration is non-dispersive or dispersive. One could argue that the net flux term (the second term) in Eq. (2) already accounts for dispersive effects, as waveform changes can occur either due to the shrinking or swelling of the wave or through the actual exchange of wave energy between neighboring wave segments caused by variations in the dispersion relation as the wave deforms during propagation.

Finally, it is worth noting that this reconciliation can be extended to waves in three-dimensional space. This extension can be achieved by applying the energy conservation law and formulating the energy budget in three dimensions. In this case, $\omega_i \propto c_{g,x} \propto c_{g,y} \propto c_{g,z}$, where the subscripts $x$, $y$, and $z$ denote the coordinate directions in three-dimensional space. Through this approach, the wave action conservation equation naturally reduces to the traditional form presented by Bretherton and Garrett ({\it 6}).

\section*{Acknowledgments}
ZW has been supported by the US National Science Foundation grants AGS-1139479, AGS-1723300, and AGS-2307335. YH was supported by the National Natural Science Foundation of China grant 42488201.

\section*{Data Availability}
This study does not involve any real-world data.

\section*{Declarations}
The authors declare that they have no conflict of interest.

\section*{Author Contributions}
Z. W. conceived the initial idea for this study and developed the arguments for each step of the proof, drawing on critical discussions with coauthors over the past 13 years. Z. W. drafted the initial version of the manuscript, and all authors contributed to its writing.

\section*{References and Notes}
\begin{enumerate}
\item F. S. Crawford, {\it Waves, Berkeley Physics Course Volume 3 \/} (MaGraw-Hill, 1968).

\item A. P. French, {\it Vibrations and Waves \/} (CRC press, 2017).

\item G. B. Whitham, {\it Linear and Nonlinear Waves \/} (Wiley \& Sons, 2011).

\item O. Bühler, {\it Waves and Mean Flows \/} (Cambridge University Press, 2014).

\item G. B. Whitham, {\it J.\ Fluid Mech.} {\bf 22}, 2 (1965).

\item F. P. Bretherton and C. J. R. Garrett, {\it Proc.\ R.\ Soc.\ A.} {\bf 302}, 1741 (1968).

\item R. Grimshaw, {\it Lecture 9- Nonlinear waves in a variable medium} \url{https://gfd.whoi.edu/wp-content/uploads/sites/18/2018/03/lecture9-roger_136584.pdf}, (2009).

\item D. Holiday, R. Resnick and J. Walker, {\it Fundamentals of Physics \/} (John Wiley \& Sons, 2013).

\item W. D. Hayes, {\it Proc.\ R.\ Soc.\ A.} {\bf 320}, 1541 (1970).

\item J. L. Lagrange, {\it Analytical Mechanics \/} (Springer Science \& Business Media, 2013)

\item D. G. Andrew and M. E. McIntyre, {\it J.\ Fluid Mech.} {\bf 89}, 4 (1978).

\item P. M. Morse and K. U. Ingard, {\it Theoretical Acoustics \/} (Princeton University Press, 1968)

\item T. H. Stix, {\it Waves in Plasmas \/} (Springer, 1992)

\item K. Aki and P. G. Richards, {\it Quantitative Seismology \/} (University Science Books, 2002)

\end{enumerate}

\end{document}



\baselineskip24pt


\maketitle 

\pagebreak


\pagebreak 
\renewcommand{\theequation}{S\arabic{equation}}
\renewcommand\thefigure{S\arabic{figure}} 

\section* {S.1 The preliminary of waves}

Since we are concerned with waves propagating in complex media that are both varying and inhomogeneously moving in space, the waves under consideration can exhibit both amplitude and frequency changes over space. Regardless of the wave system, the fundamental form of a wave can be written as
\begin{equation}
	u\left( x,t \right)=A\left( x,t \right)e^{i\theta\left( x,t \right)}
	\label{waveform}
\end{equation}
where $i$ is the imaginary unit, $u\left( x,t \right)$ is the wave field in a space-time domain, $A\left( x,t \right)$ is the amplitude of the wave field, and $\theta\left( x,t \right)$ is the phase function. It is important to note that $\theta\left( x,t \right)$ is a monotonic function of $t$ at any given location $x$, and $A\left( x,t \right)$ varies significantly more slowly than $\theta\left( x,t \right)$, that is, $ \left| \partial A / \partial x \right| \ll \left| \partial \theta / \partial x \right|$ and $ \left| \partial A / \partial t \right| \ll \left| \partial \theta / \partial t \right|$. Here, for simplicity in mathematical expression and derivation, we have adopted a one-dimensional spatial domain, but an extension to multi-dimensional space is straightforward. 

For such a wave field, the wavenumber  $k\left( x,t \right)$ and frequency  $\omega\left( x,t \right)$ (from the perspective of a spatially local observer) are defined as
\begin{equation}
	\omega\left( x,t \right)=-\frac{\partial \theta\left( x,t \right)} {\partial t} \ \ \ \mbox{and} \ \ \ k\left( x,t \right)=\frac{\partial \theta\left( x,t \right)} {\partial x}
	\label{wavenumberFrequency}
\end{equation}
Using these definitions, one can derive the wave conservation equation:
\begin{equation}
	\frac {\partial k} {\partial t} + \frac {\partial \omega} {\partial x} = 0
	\label{waveConservationEquation}
\end{equation}

The energy density of the wave field is defined as
\begin{equation}
	E\left ( x,t \right ) = u\left ( x,t \right )u^*\left ( x,t \right ) = A^2\left ( x,t \right )
	\label{energyDensity},
\end{equation}
where $u^*$ is the complex conjugate of $u$.

\section* {S.2 Conservation of locally observed frequency}

Another principle is the constancy of frequency during wave propagation. Once a wave is excited, its frequency does not change, even when the medium changes. In many standard physics books (e.g., {\it 1}), this is taken as a principle and is generally accepted, especially when electromagnetic waves pass through the interface between two different media. Intuitively, the constancy of local frequency is understandable: as the wave propagates to a new location, any imbalance between this point and the next neighboring point causes the new location to act as a source for the wave at the neighboring point. Consequently, all locations inherit the frequency of the original wave source. However, a formal proof of this constancy for waves in a continuous medium is rarely discussed, if it exists at all.

\begin{figure}[!h]
	\centering
	\includegraphics[width=0.9\linewidth]{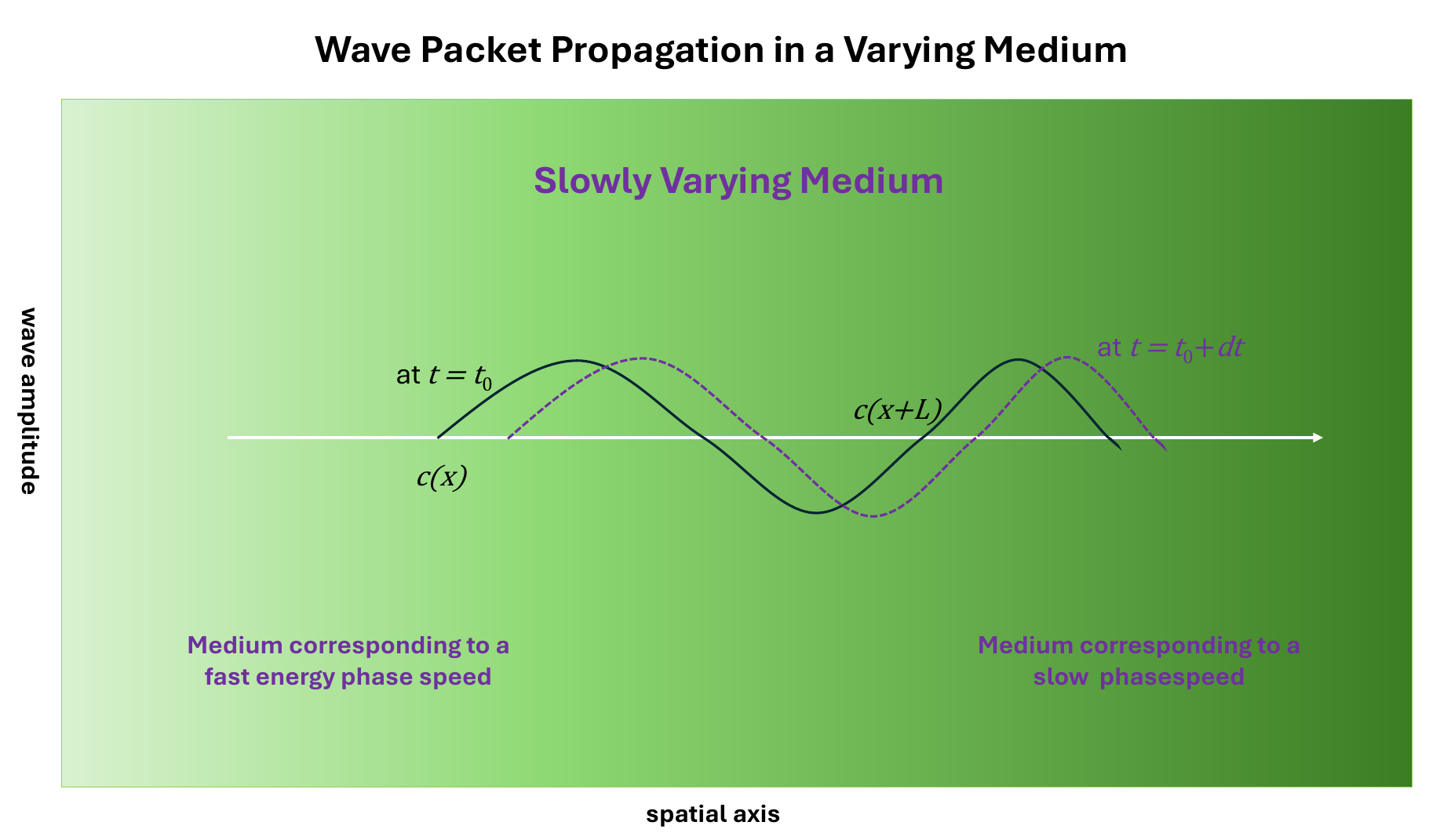}
	\caption{A schematic diagram of a wave in a continuously varying medium. The variation in background color represents the medium's variation. 
	}
	\label{Figure_A1}
\end{figure}

Here, we provide a proof of this constancy using the definition of phase velocity for the case of slowly varying medium (in contrast to a sharp medium discontinuity). We arbitrarily selected two zero-crossing points separated by one wavelength for illustration (see Fig. \ref{Figure_A1}).  At location $x$, the phase speed is $c(x)$, and at location one wave length away, $x+L$, the phase speed is $c(x)+dc/dx\cdot L$, assuming the phase speed varies slowly enough that its change can be approximated linearly. After a time $dt$, these two cross-zero points propagate to the locations $x+c(x)dt$ and $x+L+[c(x)+dc/dx\cdot L]dt$, respectively. The change of wavelength is therefore
\begin{equation}
	dL = x+L+[c(x)+dc/dx\cdot L]dt - [x+c(x)dt] - L =\frac{dc}{dx}L \frac{dx}{c}.
	\label{phasespeed_wavenumber_relation1}
\end{equation}
Since wavenumber $k(x)=2\pi/L(x)$, then we have 
\begin{equation}
	\frac{dk}{k} = - \frac{dc}{c} \ \ \Longrightarrow  \ \ kc = const \ \ \Longrightarrow \  \omega = const.
	\label{phasespeed_wavenumber_relation2}
\end{equation}
Eq. (S6) indicates that for the forward-propagating waves, the locally observed wave frequency remains constant in space. In other words, all locally observed frequencies inherit that of the incident waves.

\section* {S.3 Forms of wave action conservation equation}

\begin{figure}[!h]
	\centering
	\includegraphics[width=0.9\linewidth]{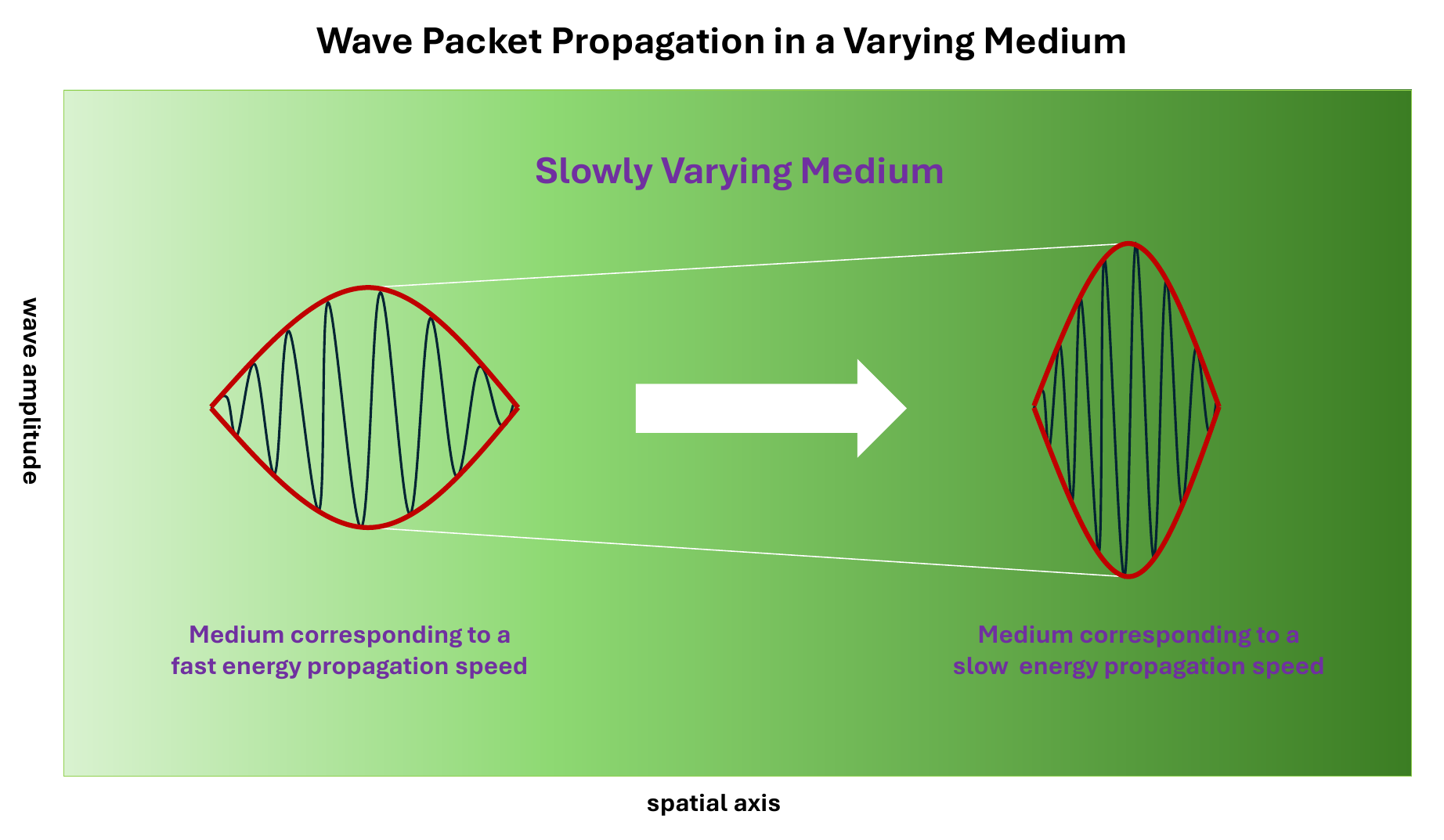}
	\caption{A schematic diagram of wave packet propagation in a continuously varying medium. The variation in background color represents the medium's variation. 
	}
	\label{Figure_A2}
\end{figure}

In the derivation of the wave action conservation equation in a varying medium, it is assumed that waves propagate as wave packets, as illustrated in Fig. \ref{Figure_A2}. A wave packet exhibits the following properties:
\begin{itemize}
	\item Each wave packet contains many wave troughs or peaks.
	\item Throughout the wave packet propagation, the number of wave peaks (or troughs) within the packet is conserved.
	\item If the medium is slowly varying, the wave packet deforms negligibly, except for proportional spatial dilations in wavelength (wavenumber) and packet amplitude shape.
	\item The wave packet propagates through the medium at the group velocity, $c_g$, which is determined by the dispersion relation of the wave system and corresponds to the speed of energy transport.
\end{itemize}

These assumptions imply that the wave packet is nondispersive (otherwise, the second assumption cannot hold). Also, changes in wavenumber, intrinsic frequency, and amplitude become noticeable only after the packet has propagated a significant distance. 

Using this wave packet assumption, Bretherton and Garrett obtained the wave action equation for waves in a spatially inhomogeneously moving medium:
\begin{equation}
	\frac {d_r} {dt} \left ( \frac {A^2} {\omega _i} \right ) + \frac {A^2} {\omega _i}  \cdot \frac {\partial c_g} {\partial x}=0
	\label{waveActionConservation_A}
\end{equation}
where $d_r/dt$ is the full temporal derivative of a quantity, corresponding to the quantity change observed by an observer moving with the quantity. $\omega_i$ is the intrinsic frequency, $c_g=c_{gi}+U\left ( x \right )$ is the total group velocity, and $c_{gi}=\partial \omega_i ⁄ \partial k$ is the intrinsic group velocity. Here the term “intrinsic” refers to measurements taken in locations where the mean flow $U \left ( x \right)$ vanishes, i.e., physical quantities measured either in a resting mean state or by observers moving with the mean flow at any location. The quantity $A^2 ⁄ \omega_i$  is referred to by Bretherton and Garrett ({\it 2}) as wave action. The equation states that changes in the wave action of a wave packet are caused by the swelling or shrinking of the wave packet in space due to the convergence or divergence of the wave packet. It should be noted that Eq. (S.7) is currently expressed in a one-dimensional spatial domain for simplicity, and its three-dimensional form can be obtained by replacing the one-dimensional group velocity divergence term with a three-dimensional one.

The definition of wave action from a variational approach is largely mathematical, and its conservation is not intuitively understandable. Since the exact mathematical expressions for kinetic energy and potential energy differ among various wave systems, the difference between them—the so-called Lagrangian—can vary significantly from one wave system to another. This raises the question of whether the wave action derived from the Lagrangian should always share the same form or if it applies only to waves in an inhomogeneously moving mean flow. To address this question, an alternative formulation of wave action conservation may provide valuable insights.

In the case of waves in an inhomogeneously moving mean flow, once a wave packet is excited, its character changes solely due to the variation of the mean flow, as noted by Hoskins and Yang ({\it 3}). Mathematically, the change in a particular characteristic, $P$, such as wavenumber $k$ or intrinsic frequency $\omega _i$, can be expressed as:
\begin{equation}
	\frac {d_rP} {dt} = \frac {\partial P} {\partial t} + c_g \frac {\partial P} {\partial x} = c_g \frac {\partial P} {\partial x} \ \ \ \mbox{and} \ \ \ \frac {\partial P} {\partial t} = 0
	\label{waveCharacter_A}.
\end{equation} 
It is noted that $\partial k/\partial t = 0$ can also be derived by combining Eqs. (S.3) and (S.6). Furthermore, the presence of a mean flow induces a Doppler shift in the frequency, expressed as:
\begin{equation}
	\omega = \omega _i + U\left( x \right)k \ \ \ \mbox{and} \ \ \ c_g = c_{gi} + U
	\label{DopplerShift_A}.
\end{equation} 

Combining Eqs. (S.3), (S.6), (S.8), and (S.9) leads to
\begin{equation}
	0 = \frac {\partial k}{\partial t} = -\frac {\partial \omega}{\partial x} = - \left( -\frac {\partial \omega_i}{\partial x} + k \frac  {\partial U}{\partial x} + U \frac {\partial k}{\partial x} \right)
	\label{usefulRelation1_A},
\end{equation}  
In the case studied by Bretherton and Garratt ({\it 2}), the changes in the medium are entirely due to its motion. For a wave packet propagating in such a medium, $c_{gi}=\partial \omega _i / \partial k = c_i = \omega _i / k = const$ and both its temporal and spatial derivatives vanish. That is,
\begin{equation}
	\partial c_{gi} / \partial t = 0 \ \ \ \mbox{and} \ \ \ \partial c_{gi} / \partial x = 0
	\label{usefulRelation2_A},
\end{equation}  
leading to the conclusion that
\begin{equation}
	-\frac {\partial c_g}{\partial x} = -\frac {\partial U}{\partial x} = \frac {1}{k} \left(\frac {\partial \omega_i}{\partial x} + U\frac {\partial k}{\partial x} \right) = \frac {c_{gi}}{\omega_i} \left(\frac{c_{gi}}{c_{gi}}\frac {\partial \omega_i}{\partial x} + \frac{U}{c_{gi}}\frac {\partial \omega_i}{\partial x} \right) = \frac{c_{gi}+U}{\omega_i} \frac{\partial \omega_i}{\partial x}
	\label{usefulRelation3_A},
\end{equation} 
i.e.,
\begin{equation}
	\frac{1}{\omega_i} \frac{\partial \omega_i}{\partial x} = -\frac{1}{c_g} \frac{\partial c_g}{\partial x}  
	\label{interRelation_A}.
\end{equation} 
Since $\omega_i$ and $c_g$ in Bretherton and Garrett's case only vary in space but not time, we have 
\begin{equation}
	\frac{1}{\omega_i}  \propto c_g
	\label{Omega_i_Cg_Relation}.
\end{equation} 
Substituting Eq. (S.14) into Eq. (S.7) yields an alternative form of the wave action conservation equation
\begin{equation}
	\frac {d_r} {dt} \left(  c_g A^2 \right ) + \left ( c_g A^2 \right )  \cdot \frac {\partial c_g} {\partial x}=0
	\label{waveActionConservation2_A}
\end{equation} 
This form of the wave action conservation equation is closely related to the apparent wave energy flux. However, $c_g$ here may better be interpreted as a dilation factor as wave propagating through a varying medium. In the main text, we will derive this relationship from fundamental physical processes and the energy conservation law, demonstrating that the apparent wave energy flux corresponds to the transmitted wave energy in a varying medium. 

\section*{S.4 Wave reflection in a discontinuous medium}

Waves propagating through a discontinuous medium are expected to undergo partial reflection at the interface between adjacent media. According to Lighthill ({\it 4}), the reflection and transmission of waves can be determined by enforcing wave continuity at the interface and ensuring zero net energy flux across it.

Let $A_I$, $A_T$, and $A_R$ represent the amplitudes of the incident, transmitted, and reflected waves, respectively. Similarly, let the corresponding group velocities be $c_{gI}$, $c_{gT}$, and $c_{gR}$. The governing equations for wave amplitudes are then given by:
\begin{equation}
	A_I - A_R = A_T,
	\label{waveAmplitudeAtInterface}
\end{equation}
and
\begin{equation}
	c_{gI}A_I^2 + c_{gR}A_R^2 = c_{gT}A_T^2.
	\label{waveTransmitted}
\end{equation}
In Eq. (S.17), the reflected portion of waves is assumed to consist of only a single wave. However, this assumption does not hold for all wave systems. For instance, in atmospheric and oceanic sciences ({\it 5, 6}), When an eastward-propagating Kelvin wave encounters a rigid eastern boundary, it reflects as a spectrum of Rossby waves, each with a distinct westward group velocity and a meridionally decaying wavy structure. These Rossby waves extend deeper into higher latitudes than the original Kelvin wave, implying that the reflected wave also has a component propagating meridionally.

At the limit of full reflection, the total reflected wave energy flux must balance that of the incident wave. In this scenario, since no waves are transmitted, Eq. (S.16) dictates that $A_R = A_I$, and the reflected group velocity $c_{gR}$ must be equal in magnitude but opposite in sign to the incident group velocity $c_{gI}$.

If we conceptually treat all reflected waves as a single, lumped wave at the interface, then the averaged group velocity of the reflected waves at the media interface must also satisfy $c_{gR}=-c_{gI}$. Assuming that the reflection dynamics remain consistent for both partial and full reflection, this result should hold generally when considering the collective effect of the reflected waves, especially for the transmitted wave that is not affected by the complexity of the reflected waves.

With the above arguments, we can derive that 
\begin{equation}
	A_R = \frac {c_{gT}-c_{gI}}{c_{gT}+c_{gI}}A_I,
	\label{waveAmplitudeReflection}
\end{equation}
and
\begin{equation}
	A_T = \frac {2c_{gI}}{c_{gT}+c_{gI}}A_I.
	\label{waveAmplitudeTransmitted}
\end{equation}
The above results align with classical theories of wave reflection at a media interface ({\it 7}). Specifically, they reinforce the principle that energy flux continuity governs wave reflection. This agreement with established wave dynamics further validates the generality of our findings.

\section*{S.5 Wave transmission in a continuously varying medium}

The waveform in Eq. (S1) represents only a forward-propagating wave and does not account for any reflected waves in a continuous medium. However, in a continuously varying medium, a portion of the forward-propagating wave can be reflected, resulting in a gradual loss of energy from the forward wave packet as it continues to propagate. In this subsection, we derive the transmission formula, building on the results established in the previous subsection.

We reinterpret Eq. (S16) for the case of a continuously varying medium. Such a medium can be approximated as a piecewise constant medium with infinitesimally small variations in its characteristics, denoted as $dc_{gI}$, similar to the approach used in calculus for integration.

For a forward-propagating wave, the transmitted portion acts as the new incident wave for the next piecewise constant segment of the medium. Since the variation in medium properties is infinitesimally small, by using Eq. (S18), the corresponding change in the incident wave's amplitude, i.e., Eq. (S16), can be expressed as
\begin{equation}
	dA_I = A_T - A_I = -A_R \approx -\frac{dc_{gI}}{2c_{gI}}A_I.
	\label{changeOFIncidentWave}
\end{equation}
Rearranging Eq. (S20) results in
\begin{equation}
	d\ln{A^2} = -\frac{dc_g}{c_{g}},
	\label{DiffChangeOFIncidentWave}
\end{equation}
Here, we omit the subscript 'I' for the forward-propagating wave. Thus, for a forward wave traveling through a varying medium, where the wave velocity changes from an initial $c_{gi}$ to a final $c_{gf}$, the change in wave energy density can be expressed as
\begin{equation}
	{A^2(c_{gf})} = A^2(c_{gi})e^{-\int_{c_{gi}}^{c_{gf}} {\frac{1}{c_g}dc_g}}
	\label{IntegChangeOFIncidentWave}.
\end{equation}

Equation (S22) takes the same form as the radiative transfer equation for absorption, provided that the integral term in this equation is interpreted as the optical depth.  

\section*{S.6 Wave action conservation derived from wave energy conservation law}

As previously mentioned, the traditional conservation law applied to propagating waves is wave action conservation, rather than wave energy conservation. Since wave action conservation is derived from variational principles and rooted in first principles, it is expected to hold. However, energy conservation is a fundamental law governing the physical world, making it difficult to conceive that it would fail for wave propagation. This apparent contradiction has remained unresolved for the past sixty years.

Here, we resolve this apparent contradiction. In the variational approach, the specific physical processes involved in wave propagation do not need to be explicitly analyzed. Instead, once the kinetic and potential energy are correctly formulated and their difference (the Lagrangian) is defined, the problem can be solved using the Euler-Lagrange equation. The fundamental principle behind this approach is that nature always evolves in the most efficient—or equivalently, the least effortful—way. Therefore, our reconciliation begins with the law of energy conservation and culminates in the wave action conservation expressed by Eq. (S15).

The process turns out to be quite simple. We follow the same procedure used to derive the continuity equation in hydrodynamics, where the conserved quantity is mass: the local temporal change in density is given by the net mass convergence divided by the local volume. Since energy, like mass, is a conserved quantity, the local temporal change in energy density ($A^2$) should similarly be given by the volume-normalized net energy flux convergence.

To make this concept easier to understand, our derivation focuses on a one-dimensional wave energy budget, as illustrated in Figure 2 of the main manuscript. In this setup, a piecewise constant medium approximates a continuous medium, similar to the concept of integration in calculus.

We consider the segment from $x-\Delta x/2$ to $x+\Delta x/2$. The net energy flux entering this segment at $x-\Delta x/2$, after deducting the reflected energy, is $c_g(x)A^2(x)$, while the energy fluxes exiting at the right side of $x+\Delta x/2$ is $c_g(x+\Delta x)A^2(x+\Delta x)$. Thus, the local temporal change of energy density can be expressed as
\begin{equation}
	\frac{\partial A^2}{\partial t} = -\frac{\partial(c_gA^2)}{\partial x}.
	\label{localEnergyDensityChangeByFlux}
\end{equation}

Additionally, due to the piecewise discontinuity at $x+\Delta x/2$, a portion of the wave energy is reflected at this boundary. This reflected energy should not be counted as part of the energy of the forward propagating wave. Based on Eq. (S19), the energy density change caused by the reflection is

\begin{align*}
	\Delta A^2 &= A^2-[\frac{2c_g(x)}{c_g(x)+c_g(x+ \Delta x)}]^2 A^2 \\
	&\approx A^2-[\frac{2c_g(x)}{2c_g(x)+ \frac{\partial c_g(x)}{\partial x} \Delta x}]^2 A^2 \\
	&\approx A^2-[1-\frac{1}{2c_g} \frac {\partial c_g(x)}{\partial x} \Delta x]^2 A^2 \\
	&\approx A^2-[1-\frac{1}{c_g} \frac {\partial c_g(x)}{\partial x} \Delta x] A^2 \\
	&= \frac{1}{c_g} \frac {\partial c_g(x)}{\partial x} \Delta x A^2. 
\end{align*}
Since $\Delta x = c_g \Delta t$, the averaged temporal  energy flux caused by the reflection within this segment is 
\begin{equation}
	\Delta A^2 = -\frac{1}{c_g}\frac { \partial c_g(x)}{\partial x} c_g\Delta t A^2 = -\frac { \partial c_g(x)}{\partial x} \Delta t A^2.
	\label{reflectedEnergyFlux}
\end{equation}
At the limit of $\Delta t$ approaches 0, we have
\begin{equation}
	\frac{\partial A^2}{\partial t} = -\frac{\partial c_g}{\partial x}A^2.
	\label{localEnergyDensityChangeByReflection}
\end{equation}
Combining both the effect of energy convergence and reflection, one obtains
\begin{equation}
	\frac{\partial A^2}{\partial t} = -\frac{\partial(c_gA^2)}{\partial x} -\frac{\partial c_g}{\partial x}A^2.
	\label{totalLocalEnergyChange}
\end{equation}
Since $c_g$ is only a function of space and not time, multiplying Eq. (S26) by $c_g$ results in
\begin{equation}
	\frac{\partial (c_gA^2)}{\partial t} + c_g\frac{\partial(c_gA^2)}{\partial x} + (c_gA^2)\cdot \frac{\partial c_g}{\partial x} = 0.
	\label{totalLocalEnergyChange}
\end{equation}
Eq. (S27) is the same as Eq. (S15); that is, we have derived wave action conservation from the energy conservation law. It is noted that this equation can further be transformed into a simpler form by multiplying $c_g$ when 
\begin{equation}
	\frac{\partial (c_g^2A^2)}{\partial t} + c_g\frac{\partial(c_g^2A^2)}{\partial x} = 0,
	\label{conservationOfFluxOfFlux}
\end{equation}
or another form
\begin{equation}
	\frac{d_r (c_g^2A^2)}{dt} = 0,
	\label{finalConservationForm1}
\end{equation}
or
\begin{equation}
	c_g^2A^2 = const.
	\label{finalConservationForm2}
\end{equation}
Eq. (S30) states that, for a one-dimensional wave, the product of the wave group velocity and the wave amplitude remains constant during wave propagation in a varying medium.

\section*{References and Notes}
\begin{enumerate}
	\item D. J. Griffiths and D. F. Schroeter, {\it Introduction to Quantum Mechanics \/} (Cambridge University Press, 2018).
	
	\item F. P. Bretherton and C. J. R. Garrett, {\it Proc.\ R.\ Soc.\ A.} {\bf 302}, 1741 (1968).
	
	\item B. J. Hoskins and G.-Y. Yang, {\it J.\ Atmos.\ Sci.} {\bf 73}, 2 (2016).
	
	\item M. J. Lighthill and J. Lighthill, {\it Waves in Fluids \/} (Cambridge University Press, 2001).
	
	\item A. J. Clarke, {\it An Introduction to the Dynamics of El Niño and the Southern Oscillation \/} (Elsevier, 2008).
	
	\item E. S. Sarachik and M. A. Cane, {\it The El Nino-Southern Oscillation Phenomenon \/} (Cambridge University Press, 2010).
	
	\item P. Malischewsky, {\it Surface Waves and Discontinuities \/} (Walter de Gruyter GmbH \& Co KG, 1987).
		
\end{enumerate}